\documentstyle[preprint,aps]{revtex}
\begin{document}
\draft
\preprint{}
\title{Increasing Potentials in Non-Abelian and
Abelian Gauge Theories}
\author{D. Singleton}
\address{Department of Physics, Virginia Commonwealth University,
1020 West Main St., Box 842000, Richmond, VA 23284-2000}
\author{A. Yoshida}
\address{Department of Physics, University of Virginia,
Charlottesville, VA 22901}
\date{\today}
\maketitle
\begin{abstract}
An exact solution for an SU(2) Yang-Mills field coupled to a
scalar field is given, which
has potentials with a linear and a Coulomb part. This
may have some physical importance since many phenomenological
QCD studies assume a linear plus Coulomb potential. Usually
the linear potential is motivated with lattice
gauge theory arguments.
Here the linear potential is an exact result of
the field equations. We also show that in the Nielsen-Olesen
Abelian model there is an exact solution in the BPS limit,
which has a Coulomb-like electromagnetic field and a
logarithmically rising scalar field. Both of these solutions
must be cut-off from above to avoid infinite field energy.
\end{abstract}
\pacs{PACS numbers:}
\newpage
\narrowtext

\section{Introduction}

In this paper we will examine some simple, exact solutions to the
field equations of both non-Abelian and Abelian gauge theories.
In both cases the gauge fields will be coupled to a scalar field
in the BPS limit \cite{prasad} \cite{bogo}
({\it i.e.} the scalar field has
zero mass and zero self coupling). We will not apply the usual
boundary conditions that the fields vanish at spatial infinity.
This means that these solutions will have infinite field
energy. Prasad and Sommerfield \cite{prasad} have given an exact,
classical solution to the SU(2) Yang-Mills-Higgs field equations
in the BPS limit, that had
non-singular fields and finite energy. Recently a new exact
solution to these field equations was discovered \cite{sing}.
This new solution was found using the analogy between
general relativity and Yang-Mills theory. It
was similiar to the Schwarzschild solution, but the character
of the spherical singularity of the ``event horizon'' was
different. Although the fields of this Yang-Mills solution
vanished at spatial infinity, the field energy was infinite
due to the singularities in the fields.

In this paper we will examine another exact solution to this
SU(2) Yang-Mills-Higgs system, which has neither fields that
vanish at infinity, nor finite energy. The time
component of the gauge fields and the scalar fields are both
found to increase linearly with distance from the origin,
while the space components of the gauge fields
have a Coulomb-like behaviour. This is of interest since
some phenomenological studies of QCD use a linear plus
Coulomb potential \cite{eichen}. Usually the linear, confining
part of the potential is motivated using lattice gauge theory
arguments \cite{wilson}. In this paper the linear potential is
an exact, analytical result of the field equations. We will also
take a look at a model with an Abelian gauge field coupled
to a scalar field. The solution for this model gives a
Coulomb-like potential for the gauge field and a rising logarithmic
scalar field. This solution suffers from both a singularity
at the origin, due to the Coulomb-like potenital, and infinite
field energy, due to the logarithmic scalar field.

We will briefly setup the field equations and simplify them
using a generalized Wu-Yang ansatz \cite{yang}. A short review of
the Schwarzschild-like classical solutions will be given so
that comparisions between these solutions can be made.

\section{Linear Potential for SU(2) Yang-Mills Theory}

The system which we consider is an SU(2) gauge field coupled
to a scalar field in the triplet representation. The Lagrangian
for this system is
\begin{equation}
\label{lag}
{\cal L} = -{1 \over 4} F_{\mu \nu} ^a F^{a \mu \nu} +
{1 \over 2} (D_{\mu} \phi ^a) (D^{\mu} \phi ^a)
\end{equation}
where the field tensor is defined in terms of the gauge fields,
$W_{\mu} ^a$, by
\begin{equation}
F_{\mu \nu} ^a = \partial _{\mu} W_{\nu} ^a - \partial _{\nu}
W_{\mu} ^a + g \epsilon ^{abc} W_{\mu} ^b W_{\nu} ^c
\end{equation}
and the covariant derivative of the scalar field is
\begin{equation}
\label{covder}
D_{\mu} \phi ^a = \partial _{\mu} \phi ^a + g \epsilon^{abc}
W_{\mu} ^b \phi ^c
\end{equation}
The Lagrangian of Eq. (\ref{lag}) is in the BPS limit
where the scalar fields' mass and self interaction are taken as zero.
In order to simplify the Euler-Lagrange equations which result from
this system one uses a generalized Wu-Yang ansatz \cite{yang}
\begin{eqnarray}
\label{wuyang}
W_i ^a &=& \epsilon _{aij} {r^j \over g r^2} [1 - K(r)]
+ \left({r_i r_a \over r^2} - \delta _{i a} \right)
{G(r) \over g r} \nonumber \\
W_0 ^a &=& {r^a \over g r^2} J(r) \nonumber \\
\phi ^a &=& {r^a \over g r^2} H(r)
\end{eqnarray}
The second term of $W_i ^a$ is usually not written down in the
Wu-Yang ansatz, and it is symmetric in its free indices as compared
to the first term which is antisymmetric. In terms of this ansatz
the Euler-Lagrange equations for this system are simplified into
the following set of four coupled, non-linear differential equations
\begin{eqnarray}
\label{difeq}
r^2 K'' &=& K (K^2 + G^2 + H^2 - J^2 - 1) \nonumber \\
r^2 G'' &=& G (K^2 + G^2 + H^2 - J^2 - 1) \nonumber \\
r^2 J'' &=& 2J (K^2 + G^2) \nonumber \\
r^2 H'' &=& 2 H (K^2 + G^2)
\end{eqnarray}
where the primes denote differentiation with respect to $r$. In
Refs. \cite{sing} and \cite{sing3} it was found that these
equations have the following solution
\begin{eqnarray}
\label{soln}
K(r) &=& {\mp cos \theta \; \; C r \over 1 \pm Cr} \nonumber \\
G(r) &=& {\mp sin \theta \; \; C r \over 1 \pm Cr} \nonumber \\
J(r) &=& {sinh \gamma \over 1 \pm Cr} \nonumber \\
H(r) &=& {cosh \gamma \over 1 \pm Cr}
\end{eqnarray}
where $C$, $\gamma$, and $\theta$ are arbitrary constants. In
\cite{sing} and \cite{sing3} only the
special case $\theta = 0$ was given. Inserting
these functions back into the expressions for the gauge and scalar
fields of Eq. (\ref{wuyang}) it is found that both the plus and minus
solutions of Eq. (\ref{soln}) have singularities at $r=0$
(plus and minus refer to the signs in the denominators
of Eq. (\ref{soln})). This
singularity is of the same kind as singularities which are found
in other classical field theory solutions ({\it e.g.} the Coulomb
solution and the Schwarzschild solution). In addition the minus
solution has a spherical singularity at $r= r_0 = 1/C$. This feature
motivated a loose comparision between this solution and the
Schwarzschild solution of general relativity.
It was speculated that the spherical barrier
at $r_0$ might give a confinement mechanism similiar to the
confinement mechanism of the true Schwarzschild solution. By
treating this solution as a background field it was shown that it
did tend to confine a scalar, color charged particle to the region
$r < r_0$ \cite{yoshida}. In addtion this confined scalar particle
behaved as a fermion due to the spin from isospin mechanism
\cite{rebbi}. However even though
the character of the singularity at $r=0$ is the
same for both the Schwarzschild solution and its Yang-Mills
counterpart, the nature of the spherical singularity is different.
In the Schwarzschild case the event horizon singularity is not
a true singularity, but is rather a coordinate singularity, which
can be removed by chosing a different coordinate system in which to
express the solution. Transforming the Schwarzschild solution to
Kruskal coordinates leaves only a singularity at $r=0$.
In the case of the minus solution given in Eq.
(\ref{soln}) the singularity at $r= 1/C$ is a true singularity
of the fields.

The new solutions to Eqs. (\ref{difeq}) are
\begin{eqnarray}
\label{nsoln}
K(r) &=& cos \theta \; \; \; \; \;  G(r) = sin \theta \nonumber \\
J(r) &=& H(r) = A r^2 + {B \over r}
\end{eqnarray}
where $A$, $B$ and $\theta$ (and therfore $K$ and $G$) are arbitrary
constants. It is interesting to note that this solution can
not be obtained from the first order Bogomolny equations
\cite{bogo}. This emphasizes the fact that although all
solutions of the Bogomolny equations will also satisfy
the Euler-Lagrange equations, the reverse is not necessarily
true. Inserting these functions into the expressions for the
gauge and scalar fields of Eq. (\ref{wuyang}) we find that the
magnitude of time component of the gauge field and
the scalar field increase linearly with $r$ as well as having
a $ 1 / r^2$ part. The space
components of the gauge fields have a Coulomb-like behaviour.
Usually, when one talks about a Coulomb plus linear potential
in QCD, this is in reference to the time component of the
gauge fields only \cite{eichen} ({\it i.e.} the time component
of the gauge field is the sum of a linear plus Coulomb term).
The linear term is thought to be the result of the non-perturbative
character of the interaction, and it is conjectured to give the
confinement property of the theory. Usually this linear term
is motivated using lattice gauge theory, but here it falls out
as an analytical result. It is obvious that this field configuration
will yield an infinite field  energy since some of the fields
do not fall off at large $r$ but rather increase as long
as $A \ne 0$. (If $A=0$ and $\theta = \pi /2$ we find that
the fields given by the new solution have the same
asymptotic behaviour as $r \rightarrow \infty$ as
our previous Schwarzschild-like solution). Also for
$\theta \ne 0$ and/or $B \ne 0$ one will have singularities at
$r=0$ for both the time and space components of the gauge fields.
Both of these features are undesirable. However if one thinks of
the solution of Eq. (\ref{nsoln}) as an isolated spin 0 ``quark''
(the scalar field in this model) then the fact that one finds
an infinite energy is what is expected from other heuristic
arguments of confinement, which contend that it should cost an
infinite amount of energy to create an isolated quark. For the
special cases when $\theta = 0$ and $B=0$ the singularities
vanish in $W_i ^a$, and $\phi ^a$ and $W_0 ^a$ respectively. (For
the case $\theta = 0$ the space component of the gauge fields
vanish altogether). The calculation of the field energy is
straight forward. Using the functions of Eq. (\ref{nsoln}) the
energy in the fields is
\begin{eqnarray}
\label{energy}
E &=& \int T^{00} d^3 x \nonumber \\
&=& {4 \pi \over g^2} \int ^{r_c} _{r_b} \left(
{J^2 \over r^2} + {(rJ' - J)^2 \over 2 r^2}
+ {H^2 \over r^2} + {(r H' - H)^2 \over 2 r^2}
\right) dr \nonumber \\
&=& {4 \pi A^2 \over g^2} (r_c ^3 - r_b ^3) -
{ 8 \pi B^2 \over g^2} \left( {1 \over r_c^3} -
{1 \over r_b^3} \right)
\end{eqnarray}
We have used the fact that $K^2 + G^2 =1$ and $K' = G' = 0$.
The integral is cut off from above because of the linearly
increasing gauge and scalar fields, and it is cut off from
below because of the singularity in $\phi ^a$ and $W_0 ^a$
at $r=0$. If one lets $r_c \rightarrow \infty$ the $A^2$
term becomes infinite due to the linearly increasing
fields. The linearly increasing fields (especially $W_0 ^a$)
are similiar to some phenomenological QCD potentials which
are thought to give the confinement property. However, these
potentials are usually motivated using lattice gauge
theory arguments, rather than being analytical results.
If one lets $r_b \rightarrow 0$
then the $B^2$ term gives an infinite field energy due
to the singular fields at $r=0$. If one takes the special
case $B = 0$ then only the linearly increasing fields make the
field energy infinite. It is interesting to note that the
Coulomb-like singularities in $W_i ^a$ apparently do not
lead to a divergence in the field energy if one integrates
down to $r=0$.

Both this new solution and the previous general relativity
inspired solutions suffer from infinite field energy unless
the integrals of the energy density are cut off. In the case of
the Schwarzschild-like solutions the infinite energy came from
integrating through the two singularities of the solution
({\it i.e.} at $r=0$ and $r=1/C$). The gauge and scalar fields of
the Schwarzschild-like solution, however, vanished rapidly as
$r \rightarrow \infty$. The present linearly increasing
solution also has singularities at $r=0$ in $W_i ^a$
if $\theta \ne 0$, and in $\phi ^a$ and $W_0 ^a$ if $B \ne 0$.
Unlike the singularities of the Schwarzschild-like
solution, the singularity in $W_i ^a$ is fairly
benign in that it does not make the integral of the energy
density diverge. The singularities in $\phi ^a$ and $W_0 ^a$
on the other hand still lead to a divergent field energy.
Thus, unless one takes the trivial case where $A = B =
\theta = 0$, the total field energy still diverges
(from the linearly increasing fields and/or from the
singularities at $r=0$), when integrated over all space.
In the special case $B=0$ the infinite
field energy comes entirely from the linearly increasing
fields, and the solution can be considered well behaved in
the sense the the Coulomb-like singularity in $W_i ^a$ at
$r=0$ does not make the energy diverge. As commented earlier, if
one views this solution as an isolated spin 0 ``quark'', then the
infinite energy fits in with the idea that it should take an
infinite amount of energy to seperate two quarks.

Finally if the present solution is used as a background field
in which to study  the motion of a test particle one finds
that the particle exhibits the spin from isospin phenomenon
\cite{rebbi} due to the antisymmetric part of $W_i ^a$.
If $\theta \ne 0$ a test scalar particle
moving in the background field of the solution will behave
as a spin 1/2 particle, while a test fermion will have integer
spin. The same thing occured when we examined a test scalar
particle moving in the background field of the Schwarzschild-like
solution \cite{yoshida}. We will present the details of a similiar
study for the present solution in an upcoming paper.

\section{Logarithmic Solution of the Nielsen-Olesen Model}

Having looked at a non-Abelian system we now turn to the somewhat
easier Abelian electrodynamics model in (2 + 1) dimensions
considered by Nielsen and Olesen \cite{olesen}. The Lagrangian
for this model looks the same as that in Eq. (\ref{lag}) with
$F_{\mu \nu} ^a \rightarrow F_{\mu \nu}$ and $(D_{\mu} \phi ^a )
(D^{\mu} \phi ^a ) \rightarrow (D_{\mu} \phi) ^{\ast} (D^{\mu}
\phi)$. The Lorentz indices only run over
$(0,1,2)$ here. The field strength tensor in this
case is
\begin{equation}
F_{\mu \nu} = \partial _{\mu} A_{\nu} - \partial _{\nu} A_{\mu}
\end{equation}
and the covariant derivative for the scalar field is now
\begin{equation}
D_{\mu} \phi = \partial _{\mu} \phi - i e A_{\mu} \phi
\end{equation}
The scalar field $\phi$ is complex and carries no group index.
Again we will take the scalar field to have no mass and no
self interaction. In order to simplify the field equations
which result from the Lagrangian we make the following ansatz
\begin{eqnarray}
\label{ansat}
A_r &=& A_0 = 0 \; \; \; \; \; A_{\theta} = A(r) \nonumber \\
\phi &=& F(r) e^{i n \theta}
\end{eqnarray}
where we are using polar coordinates $(r, \theta)$. In terms
of this ansatz the Euler-Lagrange equations of this (2+1)
Abelian gauge theory become \cite{olesen}
\begin{eqnarray}
\label{noeqn}
r F'' + F' &=& r F \left({n \over r} - e A \right) ^2
\nonumber \\
A'' + {1 \over r} A' - {1 \over r^2} A &=& \left( e^2 A
- {n e \over r} \right) F^2
\end{eqnarray}
This system of of coupled equations is simpler than the
model considered by Nielsen and Olesen who included
a mass and self interaction term for the scalar field.
If one does not require that the fields vanish as
$r \rightarrow \infty$ then a solution to Eq (\ref{noeqn})
is
\begin{equation}
\label{nosoln}
A(r) = {n \over e r} \; \; \; \; \; F(r) = F_0 \; ln (C r) + B
\end{equation}
where $F_0$, $B$ and $C$ are arbitrary constants. As with the
solution of the previous section, this solution has a field which
increases without bound as $r \rightarrow \infty$. Here
it is only the scalar field which increases, and it
increases logrithmically rather than linearly. The solution
has a Coulomb singularity at $r=0$ due to $A_{\theta}$,
and it has an infinite field energy when integrated over
all space.

\section{Conclusions}

We have given exact infinite energy solutions to two
gauge theories in the BPS limit. The first system
was an SU(2) gauge theory coupled to a scalar field in the
triplet representation. This system has been well studied
and several other infinite energy and finite energy solutions
to this model are known. In the solution given here the
scalar fields and the time component of the gauge fields
increased linearly with $r$, while the space part of the gauge
fields had a Coulomb-like behaviour. The field energy of this
solution was infinite due to the increasing fields
(and also the singularities in $\phi ^a$ and $W_0 ^a$
if $B \ne 0$), but surprisingly the Coulomb-like singularity
at $r=0$ in $W_i ^a$ did not cause the integral of
the energy density to diverge. This should be contrasted
with the Coulomb solution of electrodynamics, where the
singularity at $r=0$ does cause the integral of the energy
density to diverge. This solution has some features which
may be of interest to the study of the confinement
problem in QCD. Many phenomenological models of QCD employ
a linear plus Coulomb potential to study the spectra of
various strong interaction bound states \cite{eichen}. The
Coulomb term is thought to arise for the same reasons as
in QED, while the linear, confining term  is said to be a
result of the non-perturbative nature of the interaction.
Usually lattice gauge theory arguments are used
to motivate the linearly
confining potential. In the solution given here the
linearly increasing potential comes out as an exact result
of the classical field equations. The break
up of the linear and Coulomb parts of the solution given
here is not the same as that used in the phenomenological
studies. One could do similiar studies with the current
solution as a background field to determine if it can
reproduce some of the successes of these phenomenological
studies. In order to do this it would be necessary to give
the SU(3) version of this solution. It would be useful to
extend the solution to SU(N), which would then incorporate
the SU(3) case. As with the SU(N) generalization of
the BPS solution \cite{wilk}, and the SU(N) generalization
of the Schwarzschild-like solutions \cite{sing2}, it is
possible to extend the present solution to SU(N) by using
a maximal embedding of the SU(2) solution in SU(N).

The second solution which we considered was for the
Nielsen-Olesen model in the BPS limit. This system also
gave a solution with an increasing field. In this case it was
only the scalar field which increased, and the increase was
logarithmic rather than linear. The gauge field of this solution
was a simple Coulomb-like potential which had a singularity
at $r=0$. While the physical uses of this infinite energy
solution to the Nielsen-Olesen model are unclear, it does
share some of the characteristics of the Yang-Mills-Higgs
solution despite arising from an Abelian gauge theory.

The view taken here and in our previous paper \cite{sing}
is that some of the chacteristics of the strong interaction
(the confinement property  in particular) may be explained,
at least partially, by considering classical solutions
of the Yang-Mills system. Both the Schwarzschild-like solution
and the present solution do lead, classically, to a type of
confinement \cite{yoshida}. Whether any of these solutions
are in fact connected to the actual confinement mechanism
remains to be seen. The present solution has the advantage
over our previous Schwarzschild-like solution in
that it agrees in a loose way with the heuristic
expectations of how the confining potential for QCD
should behave. It may be that the actual confinement
mechanism of QCD is an entirely quantum effect which can
not be studied using these classical solutions. However,
given the rich structure displayed by the classical non-Abelian
gauge system, and the suggestive nature of these field
potentials, it is worthwhile to examine the possibility
that the confinement mechanism may be connected, at least
partially, to these, or possibly other undiscovered, classical
solutions.

\section{Acknowledgements} (D.S.) would like to thank
Justin O'Neill and Ruby Hutchins for help and encouragement.
Also both authors wish to acknowledge many interesting
discussions with Prof. P.K. Kabir.

\end{document}